\documentclass[aip,graphicx,jcp, reprint,amsmath,amssymb,twocolumn,usenames,dvipsnames]{revtex4-1}
\usepackage{dcolumn}
\usepackage{bm}
\usepackage{amsmath}
\usepackage{amssymb}
\usepackage{natbib}
\usepackage{graphicx}
\usepackage[usenames,dvipsnames]{color}
\draft 
\newcommand{\bra}[1]{\left\langle #1 \right|}
\newcommand{\ket}[1]{\left|#1\right\rangle}

\newcommand{\dens}[2]{\ket{#1}\bra{#2}}

\newcommand{\expt}[1]{\left\langle#1\right\rangle}
\newcommand{\abs}[1]{\left|#1\right|}
\draft 

\begin{document}


\title{A hybrid memory kernel approach for condensed phase non-adiabatic dynamics }


\author{Diptarka Hait}
\affiliation{Department of Chemistry, Massachusetts Institute of Technology, 77 Massachusetts Avenue, Cambridge, Massachusetts 02139, USA}
\affiliation{Kenneth S. Pitzer Center for Theoretical Chemistry, Department of Chemistry, University of California, Berkeley, California 94720, USA}
\author{Michael G. Mavros}
\author{Troy Van Voorhis}
\email[]{tvan@mit.edu}
\affiliation{Department of Chemistry, Massachusetts Institute of Technology, 77 Massachusetts Avenue, Cambridge, Massachusetts 02139, USA}


\date{\today}

\begin{abstract}
The spin-boson model is a simplified Hamiltonian often used to study non-adiabatic dynamics in large condensed phase systems, even though it has not been solved in a fully analytic fashion. Herein, we present an exact analytic expression for the dynamics of the spin-boson model in the infinitely slow bath limit and generalize it to approximate dynamics for faster baths. We achieve the latter by developing a hybrid approach that combines the exact slow-bath result with the popular NIBA method to generate a memory kernel that is formally exact to second order in the diabatic coupling but also contains higher-order contributions approximated from the second order term alone. This kernel has the same computational complexity as NIBA, but is found to yield dramatically superior dynamics in regimes where NIBA breaks down---such as systems with large diabatic coupling or energy bias. This indicates that this hybrid approach could be used to cheaply incorporate higher order effects into second order methods, and could potentially be generalized to develop alternate kernel resummation schemes.
\end{abstract}

\maketitle 

\section{Introduction} 
Condensed phase chemical processes are strongly influenced by the large number of solvent degrees of freedom that interact with the reacting molecules\cite{marcus1956theory,marcus1965theory,marcus1993electron,barbara1996contemporary}. Accurate modeling of condensed phase dynamics thus requires incorporation of solvent effects, but this is difficult to achieve via atomistic \textit{ab initio} techniques as the computational cost scales steeply with the number of solvent coordinates treated quantum mechanically. The generality of such atomistic approaches\cite{tully,cao1994formulation,cao1994formulation2,ben1998nonadiabatic,ben2000ab,craig2004quantum} makes them extremely useful nonetheless\cite{hammes1994proton,toniolo2003conical,toniolo2004conical,menzeleev2011direct}, but they may not represent the best tools for many problems. 

The success of Marcus theory\cite{marcus1956theory,marcus1965theory,marcus1993electron}, on the other hand, suggests that simple model systems containing a few parameters (obtained from experiment or calculations on single potential energy surfaces) could be an alternate route to study condensed phase non-adiabatic dynamics. A well known example of such a model is the spin-boson model\cite{Caldeira}, which is used quite often to study the dynamics of electron transfer\cite{barbara1996contemporary,bader1990role,xu1994coupling,evans1998photoinduced} as it reduces to the highly successful Marcus theory under appropriate limits\cite{xu1994coupling}. This suggests that it can be viewed as a dynamical generalization of Marcus theory that could be useful for studying dynamics of electron transfer and non-adiabatic molecular dynamics in general. 

However, an exact analytic solution for the population dynamics of the spin-boson model has not yet been found, despite the relative simplicity of the Hamiltonian. Various numerical approaches like Quasi-Adiabatic Propagator Path Integral (QUAPI) \cite{makarov1993tunneling,makarov1994path,topaler1994quantum,makri1995numerical,makri1995tensor,makri1995tensor2}, Multi-Configurational Time-Dependent Hartree (MC-TDH)\cite{wang2001systematic,thoss2001self}, Hierarchical Equations of Motion (HEOM)\cite{ishizaki2005quantum,tanimura2006stochastic} and Quantum Monte Carlo (QMC)\cite{qmc1,qmc2} have therefore been used to study the dynamics of the spin-boson model. Many of these approaches are formally numerically exact, but often incur significant computational cost that makes exploration of parameter space difficult. Analytic approximations like the Non-Interacting Blip Approximation (NIBA)\cite{Caldeira,chakravarty1984dynamics,aslangul1985quantum} and generalized quantum master equations\cite{muk,mavros2014resummed,kelly2016generalized} are also often used to study the dynamics of the spin-boson model, although they may also be quite computationally expensive or be effective only in certain parameter regimes.

In this paper, we present an exact analytic solution to the spin-boson model in the limit of an infinitely slow bath, and combine this result with NIBA to generate a hybrid kernel for the dynamics of general baths with arbitrary spectral densities.  This hybrid kernel approximation has the same computational complexity as NIBA, but is found to yield much superior results in regimes where the latter fails, being approximately as effective as a fourth order resummed memory kernel approach for all cases tested. In particular, it is found to give quite good performance in problematic cases involving low temperatures, large diabatic coupling and large energy biases, where NIBA is known to give qualitatively incorrect behavior. This indicates that the hybrid approach is a promising alternative to naive NIBA, as it has the same computational complexity but yields superior results in all regimes tested. {Consequently, it can be used to cheaply explore a large parameter space for screening purposes over a more expensive method like MC-TDH. The developed approach is also able to approximate dynamics from second order information alone, suggesting that it can potentially be generalized to develop a suite of resummation methods that can be applied to problems more general than the spin-boson model.}

\section{The spin-boson model}
The spin-boson model\cite{Caldeira} is one of the simplest models for open quantum systems, describing the interaction between a two level system and a bath of harmonic oscillators. The model assumes that the diabatic coupling $V$ between the two system states is constant and the system-bath coupling is linear in the bath coordinates. There may however be a non-zero intrinsic energy bias $\epsilon$ between the two system states. 
 
Therefore, the Hamiltonian $\mathbf{H}_{SB}$, in the basis of orthogonal diabatic states $\ket{1}=\begin{pmatrix}
1\\0
\end{pmatrix}$ and $\ket{2}=\begin{pmatrix}
0\\1
\end{pmatrix}$, is given by:
\begin{align}
\mathbf{H}_{SB}&=\begin{pmatrix}
\mathbf{h_1}&V\\V&\mathbf{h_2}
\end{pmatrix}\label{hamiltonian}\\
\mathbf{h_1}&=-\dfrac{\epsilon}{2}+\displaystyle\sum\limits_i\left(\dfrac{p_{i}^2}{2m_i}+\dfrac{1}{2}m_i\omega^2_ix_i^2+c_i x_i\right)\label{h1def}\\
\mathbf{h_2}&=\phantom{-}\dfrac{\epsilon}{2}+\displaystyle\sum\limits_i\left(\dfrac{p_{i}^2}{2m_i}+\dfrac{1}{2}m_i\omega^2_ix_i^2-c_i x_i\right)\label{h2def}
\end{align}
The harmonic bath has a set of modes $\{x_i\}$ with frequencies $\{\omega_i\}$ and masses $\{m_i\}$. These modes interact linearly with the system states $\ket{1}$ and $\ket{2}$ via the coupling constants $\{c_i\}$. Consequently, $\mathbf{h_{\{1,2\}}}$ are compact descriptions for the bath Hamiltonians associated with states $\ket{1}$ and $\ket{2}$. 

There are typically a large number of bath modes present but we are chiefly interested in the diabatic populations and not the minutiae of individual bath coordinates. It is therefore reasonable to treat the bath oscillators as a continuum via a spectral density $J(\omega)$ of the form\cite{nitzan2006chemical}:
\begin{align}
J(\omega)=\displaystyle\sum\limits_{i=1}^N\dfrac{\pi c_i^2}{2m_i\omega_i}\delta(\omega-\omega_i)\label{jdef}
\end{align} 
It is known that the dynamics of the spin-boson model is completely specified by $V,\epsilon$ and $J(\omega)$\cite{chakravarty1984dynamics,leggett1987dynamics,nitzan2006chemical}-though an actual closed form expression remains elusive.

We also note that the spin-boson model is extremely simplified and neglects many physical effects (non-Condon effects\cite{izmaylov2011nonequilibrium,mavhait}, Duschinski rotations\cite{tang2003effects} and bath anharmonicity, to name a few) that are relevant in many condensed phase processes of interest. However, the relative difficulty in solving even this simplified Hamiltonian makes it reasonable to use the spin-boson model as a starting point for method development and then generalize to more complex models.  

\section{Memory Kernels}
Memory kernels\cite{nakajima,zwanzig1,zwanzig2,zwanzig3} represent one possible approach for studying the spin-boson and related models. 
We employ a generalized version of the memory kernel formalism of Sparpaglione and Mukamel\cite{muk} in this paper, where the memory kernels $K_{11/22}(t,t_1)$ control the dynamics of the population $p_{i}(t)$ in the state $\ket{i}$ in the following manner:
\begin{align}
\dfrac{dp_1}{dt}&=\displaystyle\int\limits_{0}^t\left(K_{22}(t,t_1)p_2(t_1)-K_{11}(t,t_1)p_1(t_1)\right)\,dt_1 \label{sm1}\\
\dfrac{dp_2}{dt}&=\displaystyle\int\limits_{0}^t\left(K_{11}(t,t_1)p_1(t_1)-K_{22}(t,t_1)p_2(t_1)\right)\,dt_1 \label{sm2}
\end{align}

However, the only independent quantity here is the difference $p_1(t)-p_2(t)$ as $p_1(t)+p_2(t)=1$ for all times. 
This difference is termed as $\expt{\sigma_z}(t)$ since $p_1(t)-p_2(t)=\mathbf{Tr}\left[\bm{\rho}(t)\left(\dens{1}{1}-\dens{2}{2}\right)\right]=\mathbf{Tr}\left[\bm{\rho}(t)\sigma_z\right]$, where $\bm{\rho}(t)$ is the time-dependent density matrix for the whole system. We thus actually just have a single equation:
\begin{align}
\dfrac{d\expt{\sigma_z}}{dt}&=-\displaystyle\int\limits_{0}^t\left(K_{-}(t,t_1)+K_{+}(t,t_1)\expt{\sigma_z}(t_1)\right)\,dt_1 \label{sm3}
\end{align}
where $K_{\pm}(t,t_1)=K_{11}(t,t_1)\pm K_{22}(t,t_1)$.

Time-dependent perturbation theory allows us to expand the population difference $\expt{\sigma_z}(t)$ and the kernels $K_{\pm}(t,t_1)$ into:
\begin{align}
\expt{\sigma_{z}}(t)&=\displaystyle\sum\limits_{m=0}^\infty V^m\expt{\sigma_{z}}^{(m)}(t)\\
K_{\pm}(t,t_1)&=\displaystyle\sum\limits_{m=0}^\infty V^mK^{(m)}_{\pm}(t,t_1)
\end{align}
where $\expt{\sigma_{z}}^{(m)}(t)$ and $K^{(m)}_{\pm}(t,t_1)$ can be found via the procedure described in Appendix A. 

The popular non-interacting blip approximation (NIBA) \cite{chakravarty1984dynamics,leggett1987dynamics,aslangul1985quantum} approximates $K$ by only using the second-order term (i.e. $K\approx V^2 K^{(2)}$), yielding quite accurate dynamics in the small $V$ (`outer sphere') regime and reducing to Marcus theory for slow bath frequencies at high temperatures\cite{xu1994coupling}. It is expected that the dynamics obtained from kernels incorporating higher order terms will be even more accurate, but would be much more computationally expensive to obtain as kernels accurate to $2n$th order in $V$ require evaluation of $2n-1$ dimensional oscillatory integrals. Therefore, alternative approaches that can obtain reasonable approximations of higher order terms from only second-order terms like $\expt{\sigma_{z}}^{(2)}$ are desirable as they would incur a much lower computational cost.  

\subsection{Initial conditions}
The rest of the paper will assume that the initial density matrix $\bm{\rho}(0)=p_1(0)\dens{1}{1}\otimes \rho_{B}+p_2(0)\dens{2}{2}\otimes\rho_{B'}$, which causes odd order power series terms $\expt{\sigma_z}^{(2k+1)}(t)$ and $K_{\pm}^{(2k+1)}(t)$ to be $0$. We will perform relevant derivations with the even simpler initial condition of $\bm{\rho}(0)=\dens{1}{1}\otimes \rho_B\implies \expt{\sigma_z}(0)=1$ for mathematical simplicity, as the expressions derived from this form can be trivially generalized to any initial separable diabatic density matrix due to linearity of time evolution. In particular, we would like to emphasize that Eqn \ref{gsb1} and beyond are true for all $\bm{\rho}(0)=p_1(0)\dens{1}{1}\otimes \rho_{B}+p_2(0)\dens{2}{2}\otimes\rho_{B'}$. 

The initial bath density matrix $\rho_B$ is determined by the nature of the dynamical process of interest. Ground state electron transfer processes often use equilibrium $\rho_{B}=\dfrac{e^{-\beta \mathbf{h_1}}}{\mathbf{Tr}\left[e^{-\beta \mathbf{h_1}}\right]}$ ($\beta$ being the inverse temperature $\dfrac{1}{k_BT}$) as this describes a scenario where the bath modes are in thermal equilibrium with the reduced Hamiltonian $\mathbf{h_1}$ associated with the state $\ket{1}$ (which has all the population). On the other hand, non-equilibrium $\rho_{B}$ of the type $\dfrac{e^{-\beta \mathbf{h_2}}}{\mathbf{Tr}\left[e^{-\beta \mathbf{h_2}}\right]}$ can be useful for photochemistry, as this corresponds to a situation where the population was purely in $\ket{2}$ with the bath in thermal equilibrium, before an excitation at $t=0^-$ led transfer of all the diabatic population to $\ket{1}$ without giving the bath modes an opportunity to relax accordingly, due to difference in timescales. We call this initial condition the ``photochemical" non-equilibrium condition throughout the rest of the paper for convenience, although it need not correspond to experimentally observable photochemistry for all systems. Other non-equilibrium initial conditions may also make physical sense for different processes. Unless specified otherwise, all our expressions are valid for any {$\rho_B=\dfrac{e^{-\beta \mathbf{h_3}}}{\mathbf{Tr}\left[e^{-\beta \mathbf{h_3}}\right]}$} where $\mathbf{h_3}=\mathbf{h_0}+\displaystyle\sum\limits_i d_ix_i$ ($\{d_i\}$ can be any arbitrary real number) is a bath Hamiltonian with the same frequencies and modes as $\mathbf{h_{1/2}}$, but with any arbitrary set of equilibrium positions. 

\subsection{Power series terms}
Evaluation of the non-zero even order terms in the power series of $\expt{\sigma_z}(t)$ involves tracing over bath modes. Time-dependent perturbation theory reveals that the power series coefficients $\expt{\sigma_z}^{(2n)}(t)$ for $\bm{\rho}(0)=\dens{1}{1}\otimes \rho_B$ are completely specified by integrals of traces $f_{2n}(t_1,t_2,t_3\ldots t_{2n})$ where:
\begin{align}
f_{2n}(t_1,t_2,t_3\ldots t_{2n})&={\mathbf{Tr}\left[O(t_1)O^\dagger(t_2)O(t_3)\ldots O^\dagger(t_{2n})\rho_B\right]}\\
O(t)&=e^{i\mathbf{h_1}t}e^{-i\mathbf{h_2}t}
\end{align} 
For instance, we have:
\begin{align}
\dfrac{d}{dt}\expt{\sigma_z}^{(2)}(t)&=-4\displaystyle\int\limits_0^{t}\mathbf{Re}\left[f_2(t,t_1)\right]\,dt_1 \label{k2niba}
\end{align}
Similarly, $\expt{\sigma_z}^{(4)}(t)$ is expressed in terms of integrals of $f_4$, $\expt{\sigma_z}^{(6)}(t)$ by integrals of $f_6$ and so on. 
\subsection{Non-Interacting Blip Approximation}
Going up to second order alone, we discover that $K_{11}^{(2)}(t,t_1)=2\mathbf{Re}\left[f_2(t,t_1)\right]$ is consistent with Eqn. \ref{k2niba}. {This expression is similar to the non-equilibrium golden rule in Ref [\onlinecite{coalson1994nonequilibrium}], although the formalism employed therein is convolution free unlike Eqns \ref{sm1}-\ref{sm2}}. For the case of $\rho_{B}=\dfrac{e^{-\beta \mathbf{h_1}}}{\mathbf{Tr}\left[e^{-\beta \mathbf{h_1}}\right]}$ (thermal equilibrium initial conditions), this reduces to the well-known Non-Interacting Blip Approximation (NIBA)\cite{Caldeira,chakravarty1984dynamics,aslangul1985quantum}. The second-order kernels $K^{(2)}_{11/22}$ are then given by\cite{mavros2014resummed}:
\small
 \begin{align}
 K_{11}^{(2)}(t,t_1)&=2e^{-Q'(t-t_1)}\cos\left(Q''(t-t_1)+\epsilon \left(t-t_1\right)\right)\\
 K_{22}^{(2)}(t,t_1)&=2e^{-Q'(t-t_1)}\cos\left(Q''(t-t_1)-\epsilon \left(t-t_1\right)\right)\\
 Q' (t)&=\dfrac{4}{\pi}\displaystyle\int\limits_0^\infty \dfrac{J(\omega)}{\omega^2}\left(1-\cos \omega t\right)\coth\dfrac{\beta\omega}{2}\,d\omega\\
 Q'' (t)&=\dfrac{4}{\pi}\displaystyle\int\limits_0^\infty \dfrac{J(\omega)}{\omega^2}\sin \omega t\,d\omega
 \end{align}
 \normalsize
 where $J(\omega)$ is the spectral density defined earlier in Eqn. \ref{jdef}. It can quickly be noted that these kernels only depend on the time difference $t-t_1$ and, as such, we can simply replace the function of two variables $K_{11/22}^{(2)}(t,t_1)$ with the univariate $ K_{11/22}^{(2)}(t-t_1)$. The resulting kernel $K_{11/22}(t)\approx V^2K_{11/22}^{(2)}(t)$ is termed as the NIBA kernel. This approach is expected to yield decent dynamics for small $V$, as it is accurate to the lowest non-zero order for this model. 
 
 For non-equilibrium initial conditions however, the traces involved are not time-translationally invariant as they depend on both $(t,t_1)$ and not just the difference $t-t_1$, unlike the equilibrium case. As an example, this generalization for the case of non-equilibrium initial $\rho_B=\dfrac{e^{-\beta \mathbf{h_2}}}{\mathbf{Tr}\left[e^{-\beta \mathbf{h_2}}\right]}$results in kernels\cite{mavhait}:
 \begin{align}
 K_{11}^{(2)}(t,t_1)&=2e^{-Q'(t-t_1)}\cos\left(\phi(t,t_1)+\epsilon \left(t-t_1\right)\right)\\
 K_{22}^{(2)}(t,t_1)&=2e^{-Q'(t-t_1)}\cos\left(\phi(t,t_1)-\epsilon \left(t-t_1\right)\right)\\
 \phi(t,t_1)&=Q''(t-t_1)+2Q''(t_1)-2Q''(t)
 \end{align}
 which are very similar to the equilibrium kernels, aside from the subtle difference stemming from the phase $\phi(t,t_1)$ not being time-translationally invariant. The resulting kernels  $K_{11/22}(t,t_1)\approx V^2K_{11/22}^{(2)}(t,t_1)$ cannot strictly be called NIBA kernels, but this is effectively a natural extension of the idea behind NIBA, and is used in its place for non-equilibrium initial conditions. {Similar expressions had been derived earlier for both memory kernel\cite{thoss2001self} and non-equilibrium golden rule studies\cite{izmaylov2011nonequilibrium}.} 
 \subsection{Resummed Kernels} 
  NIBA is only correct to second order, and performs quite poorly for systems with large diabatic coupling, large energy bias or low temperature. The natural step forward seems to be adding fourth or sixth order terms to the approximate kernel. However, simply stopping there might be problematic as finite truncations of perturbation theory series are often not convergent\cite{dyson1952divergence}. Resummation techniques help ameliorate this complication by approximating the unknown higher order terms from the known low order terms. The recipe for memory kernel resummation has been outlined in Refs [\onlinecite{mavros2014resummed},\onlinecite{wu}] and other places, and we will not discuss the details here. We will merely note that this approach yields significantly superior results compared to NIBA\cite{mavros2014resummed,wu,hsing-ta} but this increased accuracy comes with the much higher computational cost of evaluating $K^{(4)}$ or other higher order kernels exactly.
\section{Trace relations}
Closed-form expressions for the traces $f_{2n}$ for $n\ge1$ are necessary for calculating $\expt{\sigma_z}^{(2n)}$ or $K^{(2n)}$. We discovered that all traces $f_{2n}$ can be expressed in terms of the second order trace $f_2$ alone for \textit{any} integer $n$ and \textit{any} choice of the spectral density; if the initial bath density matrix $\rho_B$ equals $\dfrac{e^{-\beta \mathbf{h_3}}}{\mathbf{Tr}\left[e^{-\beta \mathbf{h_3}}\right]}$ for $\mathbf{h_3}=\mathbf{h_1}+\displaystyle\sum\limits_i d_ix_i$ ($\{d_i\}$ can be any arbitrary set of real numbers of cardinality $N$). Specifically, we have:
\begin{align}
\ln f_{2n}(t_1,t_2,t_3\ldots t_{2n})&=\displaystyle\sum\limits_{i=1}^{2n-1}\displaystyle\sum\limits_{j=i+1}^{2n}  \left(-1\right)^{i+j+1}\ln f_2(t_i,t_j)\label{tracefactor}
\end{align}
For example:
\begin{align}
f_4(t_1,\ldots t_4)&=\dfrac{f_2(t_1,t_2)f_2(t_2,t_3)f_2(t_3,t_4)f_2(t_1,t_4)}{f_2(t_1,t_3)f_2(t_2,t_4)}
\end{align}
and similarly for $f_6,f_8$ etc.

Essentially, an initial bath density matrix corresponding to a thermal state of any harmonic bath $\mathbf{h_3}$ with same frequencies and modes as $\mathbf{h_{1/2}}$ is sufficient to satisfy the trace factorization relationship Eqn. \ref{tracefactor} (irrespective of the equilibrium positions of the bath modes in $\mathbf{h_3}$) for any bath spectral density. {Similar relationships had been presented earlier for equilibrium initial conditions of $\mathbf{h_3}=\mathbf{h_1}$\cite{skinner1986optical,leggett1987dynamics} but we believe that the case of general $\mathbf{h_3}$ has not previously been reported}. This relationship has been analytically verified to only 12th order for $\mathbf{h_3}=\mathbf{h_1}$ (equilibrium initial conditions) and 6th order for $\mathbf{h_3}=\mathbf{h_2}$ (``photochemical" non-equilibrium initial conditions) due to the rather steep memory cost for the evaluation of the integrals needed to find the analytic expressions with Mathematica\cite{wolfram1999mathematica}. It has however been verified numerically up to $f_{60}$ for multiple $\mathbf{h_3}$ resulting from random $\{d_i\}$ over a large number of randomly selected time indices $\{t_i\}$, indicating that it is correct to 60th order at least, and very likely beyond as well (though as of now, we do not possess a proof for this conjecture). This result is remarkable because it implies that for \textit{any} spectral density, the dynamics is a function of $f_2$ alone. That is to say that, in principle, one can predict all of the dynamics using only $f_2(t_1,t_2)$ as an input.  In analogy with time density functional theory \cite{runge1984density} (in which the dynamics of a many particle system is written as a functional of the one particle density alone), one can envision a $f_2(t_1,t_2)$ functional theory - in which the dynamics are directly predicted by some functional of $f_2$ alone.  NIBA is one such theory, but the result above suggests that the exact result can in principle be constructed from $f_2$ alone. 

 \begin{figure*}[htb!]
 	\vspace{-10pt}
 	\begin{minipage}{0.48\textwidth}
 		\centering
 		(a)\includegraphics[width=3.2in]{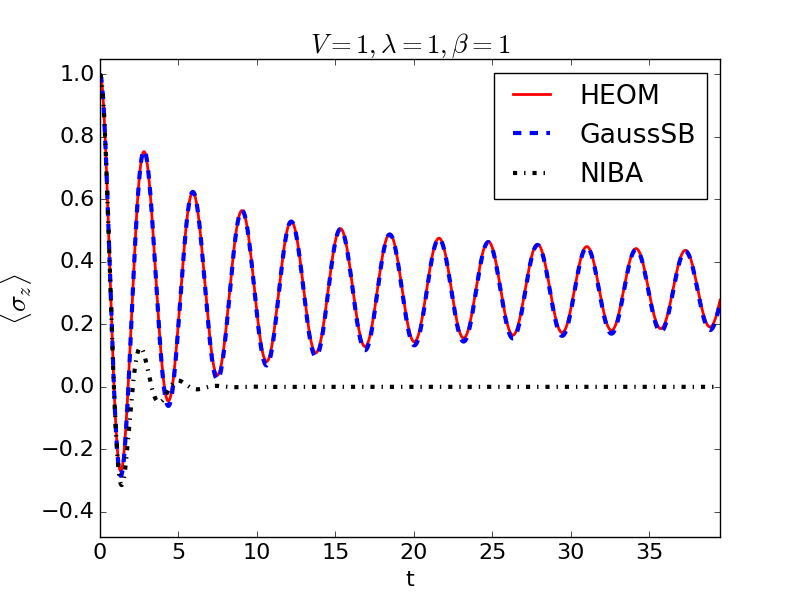}
 		\label{fig:vbl1}
 		\vspace{-5pt}
 	\end{minipage}
 	 	\begin{minipage}{0.48\textwidth}
 	 		\centering
 	 		\vspace{-3pt}
 	 		{(b)}\includegraphics[width=3.2in]{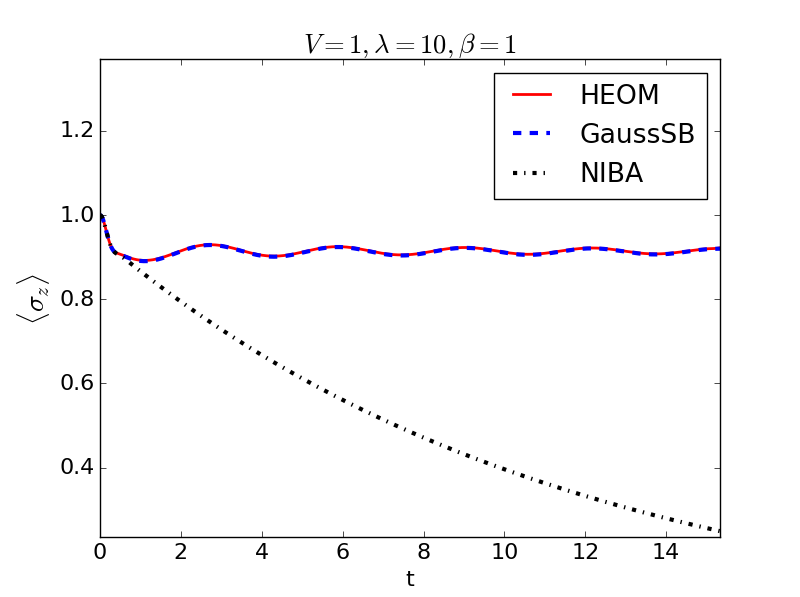}
 	 		\label{fig:v3}
 	 	\end{minipage}
 	\begin{minipage}{0.48\textwidth}
 		\centering
 		\vspace{-3pt}
 		(c)\includegraphics[width=3.2in]{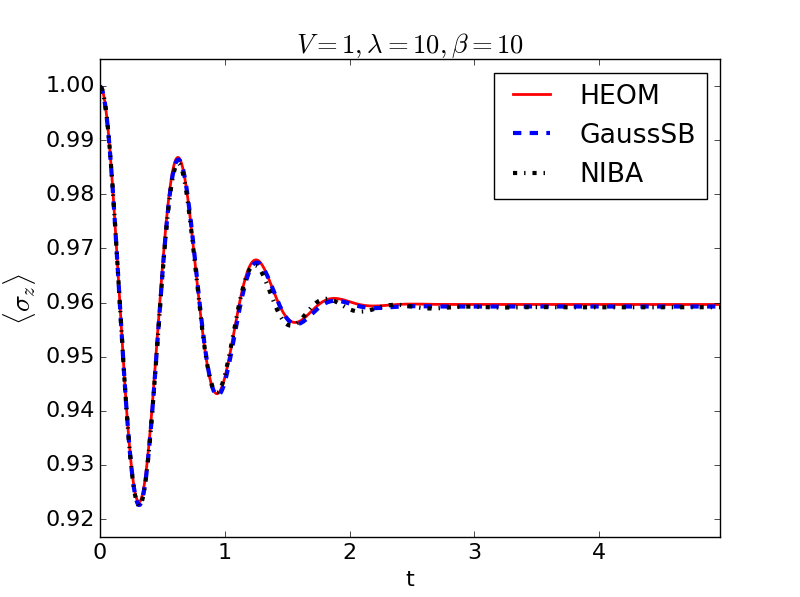}
 		\label{fig:vbl2}
 	\end{minipage}
 	\begin{minipage}{0.48\textwidth}
 		\centering
 		{(d)}\includegraphics[width=3.2in]{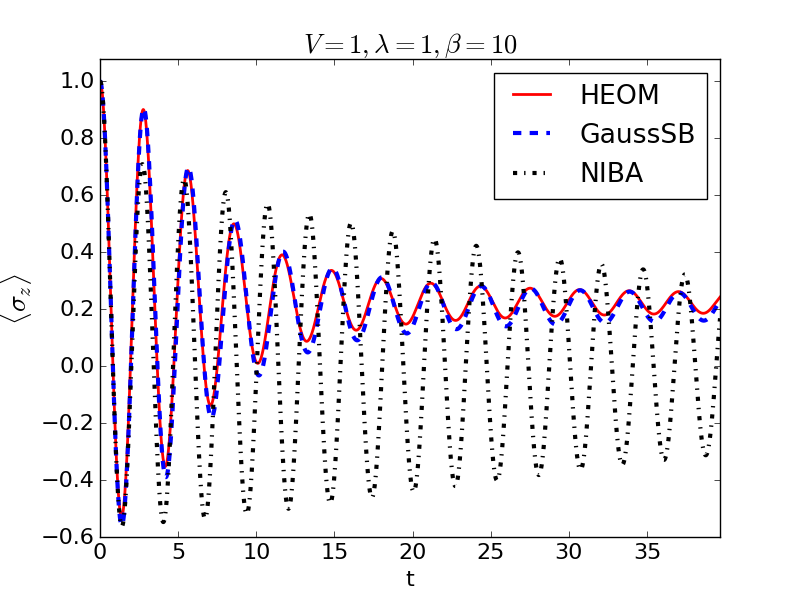}
 		\label{fig:v0}
 		\vspace{-5pt}
 	\end{minipage}
 	\vspace{-10pt}
 	\caption{Comparison between GaussSB, NIBA and HEOM at $V=1,\omega_c=10^{-4}$ and $\epsilon=0$.}
 	\label{fig2}
 	\vspace{-10pt}
 \end{figure*} 
 
\section{Case of the {Slow bath}}
Though Eqn. \ref{tracefactor} is quite useful in finding all the perturbation theory traces, it does not eliminate the need to numerically integrate $f_{2n}$ over the $2n$ time indices in order to find $\expt{\sigma_z}^{(2n)}(t)$. This quickly becomes intractable due to the grid size growing exponentially with $n$, creating a practical upper limit for applying naive perturbation theory even though we can find traces to arbitrary order $2n$ rather simply via Eqn. \ref{tracefactor}.

However at short times, we may expand $\ln f_2(t_1,t_2)$ as a Taylor series about $t_1=t_2=0$ to obtain:
\begin{align}
\ln f_2(t_1,t_2)&=ib(t_1-t_2)-a(t_1-t_2)^2+\ldots
\end{align}
where
\begin{align}
b&=\mathbf{Tr}\left[(\mathbf{h_1}-\mathbf{h_2})\rho_B\right]\\
a&=\dfrac{\mathbf{Tr}\left[(\mathbf{h_1}-\mathbf{h_2})^2\rho_B\right]-\left(\mathbf{Tr}\left[(\mathbf{h_1}-\mathbf{h_2})\rho_B\right]\right)^2}{2}
\end{align}
Higher order terms can be neglected when $\omega t_{1,2}\ll 1$, where $\omega$ is the characteristic response frequency of the system. Taking this short-time/slow-bath approximation and defining a Gaussian function $g(t)=e^{-at^2+ibt}$, we obtain:
\begin{align}
f_2(t_1,t_2)&\approx g(t_1-t_2)\\
f_{2n}(t_1,t_2\ldots t_{2n})&\approx g(t_1-t_2\ldots +t_{2n-1}-t_{2n})
\end{align}
which enables us to approximate all these traces as Gaussian functions that can be analytically integrated. Carrying out the integrations over the time indices $t_i$ and summing over all orders in $V$ (in the manner described in Appendix C), we find that:{\small 
\begin{align}
&\dot{\expt{\sigma_z}}(t)=V^2\dot{\expt{\sigma_z}}^{(2)}(t)-2V^3\displaystyle\int\limits_0^t J_1(2Vt_1)\dot{\expt{\sigma_z}}^{(2)}\left(\sqrt{t^2-t_1^2}\right)\,dt_1\label{gsb1}\\
&\dot{\expt{\sigma_z}}^{(2)}(t)=-\mathbf{Re}\left[\frac{2\sqrt{\pi } e^{-\frac{b^2}{4 a}} \text{erf}\left(\frac{2 a t+ib}{2 \sqrt{a}}\right)}{ \sqrt{a}}\right]\label{gsb2}
\end{align}}
where $J_1$ is a Bessel function of the first kind. We note that Eqn \ref{gsb1} holds for any initial density matrix of the diagonal form $p_1(0)\dens{1}{1}\otimes \rho_{B}+p_2(0)\dens{2}{2}\otimes\rho_{B'}$, courtesy the linearity of time evolution. 

For equilibrium initial conditions we have $
 b=\epsilon+\lambda$ and $
 a=\dfrac{2}{\pi}\displaystyle\int\limits_0^\infty J(\omega)\cosh \dfrac{\beta\omega}{2}\,d\omega
 $;
 where $\lambda=\dfrac{4}{\pi}\displaystyle\int\limits_0^\infty \dfrac{J(\omega)}{\omega}\,d\omega$ is the Marcusian reorganization energy. $a\approx \dfrac{\lambda}{\beta}$ in the limit of frequencies $\{\omega_i\}\to 0$. The expression for $\expt{\sigma_z}(t)$ obtained with these values for $a,b$ appears to be consistent with the results obtained for a slow bath by Hornbach and Dakhnovski\cite{hornbach1999electron}, which makes us more confident in overall accuracy of our result and the applicability to general initial conditions beyond the equilibrium case. The ``photochemical" non-equilibrium initial condition, on the other hand, yields the same $a$, but $b=\epsilon-\lambda$ instead, indicating that for slow-baths at least, equilibrium and ``photochemical" conditions lead to same dynamics if $\epsilon=0$. 
 
We can also substitute the exact $\dot{\expt{\sigma_z}}^{(2)}(t)=-4\displaystyle\int\limits_0^t\mathbf{Re}\left[f_2(t,t_1)\right]\,dt_1$ in Eqn \ref{gsb1} instead of the Gaussian approximate form given in Eqn \ref{gsb2}. This is done in the hope of generalizing Eqn \ref{gsb1} to baths with non-Gaussian $f_2$ as the substitution makes Eqn \ref{gsb1} exact to order $V^2$ for \textit{any} set of parameters. We call this 
version of Eqn \ref{gsb1} GaussSB as it is exact for spin-boson problems with Gaussian $f_2$.


\begin{figure*}[htb!]
	\centering
	
	\begin{minipage}{0.48\textwidth}
		\centering
		(a)\includegraphics[width=3.2 in]{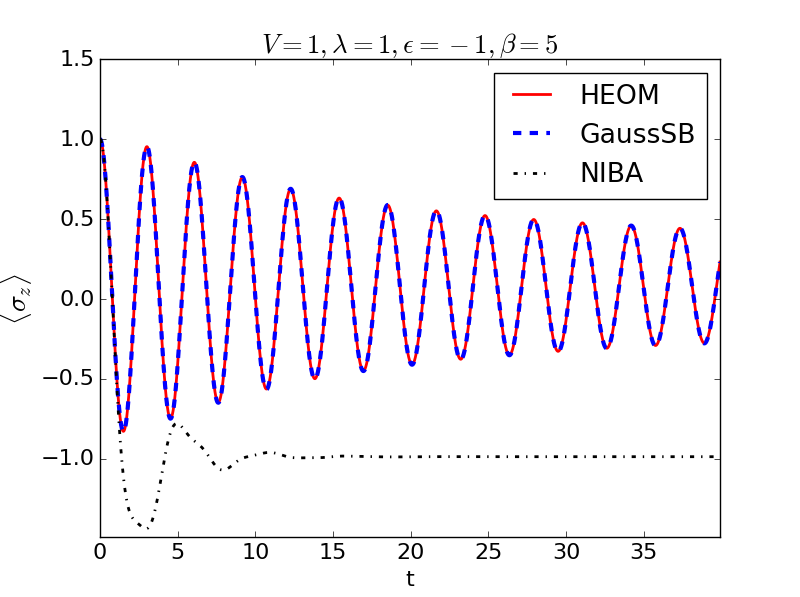}
		\label{fig:bm}
		\vspace{-10pt}
	\end{minipage}
	\begin{minipage}{0.48\textwidth}
		\centering
		(b)\includegraphics[width=3.2 in ]{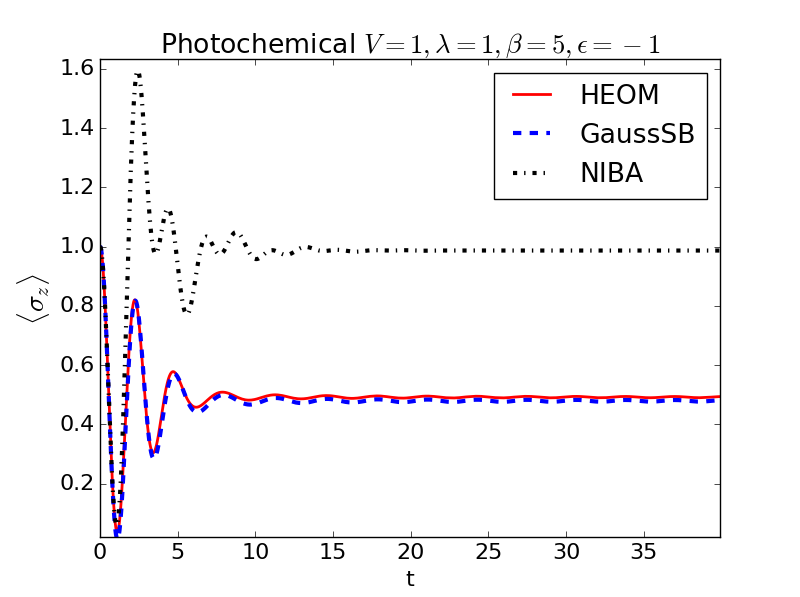}
		\label{fig:phot}
		\vspace{-10pt}
	\end{minipage}
	\caption{Comparison between GaussSB, NIBA and HEOM at $\omega_c=10^{-4}, V=1,\lambda=1,\beta=5$ and $\epsilon=-1$ for both equilibrium and ``photochemical" non-equilibrium initial conditions. }
	\label{bias}
\end{figure*}

\section{Benchmarking Accuracy of GaussSB}
The lack of a rigorous proof for Eqn. \ref{tracefactor} raises some well warranted questions about the accuracy of GaussSB even in the Gaussian limit, leading us to compare it against HEOM\cite{ishizaki2005quantum,tanimura2006stochastic}---a method known to be numerically exact for $J(\omega)=\dfrac{\lambda}{2}\left(\dfrac{\omega\omega_c}{\omega^2+\omega_c^2}\right)$ if a sufficiently large number of Matsubara frequencies are employed and a high hierarchy depth used for calculations. This particular form of $J(\omega)$ appears to be suboptimal for GaussSB on account of the long, slowly decaying high-frequency tail, but we nonetheless stick with it for convenience in benchmarking. We also compared GaussSB to NIBA in order to determine the differences between the two, since they both attempt to approximate the dynamics with the second order term $\expt{\sigma_z}^{(2)}(t)$ alone. Comparisons were done in both the slow bath limit (where $f_2$ is Gaussian and GaussSB is expected to be exact) and the fast bath limit (where $f_2$ is decidedly non-Gaussian and GaussSB is likely to fail).

HEOM calculations were done with the PHI code\cite{strümpfer2012open} while both GaussSB and NIBA were implemented in C++ employing the error function implementation by Johnson \cite{johnson2012faddeeva} and the GSL implementation of Quadpack integration routines \cite{gough2009gnu,piessens2012quadpack}. 
	
\subsection{Slow Baths} 
GaussSB is expected to be valid in the $\omega_c\to 0$ limit (slow bath limit) and so $\omega_c$ was set to $10^{-4}$ for all calculations in this subsection. We also set $V=1$ throughout in order to fix the timescale of oscillations in diabatic populations. We first checked for cases without bias ($\epsilon=0$) over a wide range of temperatures and reorganization energies ($0.1\le \lambda,\beta\le 10$). Only equilibrium initial conditions were tested, as ``photochemical" initial conditions give the same Gaussian parameters $a,b$ for cases where $\epsilon=0$, and thus should yield same dynamics. Four different behaviors were observed in this regime and these are depicted in Fig \ref{fig2}. There is visual agreement between GaussSB and HEOM in each case presented (as well as in many other cases tested that resulted in qualitatively similar dynamics), while NIBA often even fails to reproduce the qualitative behavior. These tests are consistent with our claim that GaussSB is exact in the slow bath limit. 

We next consider cases with bias, where equilibrium and ``photochemical" dynamics are expected to give different results. Two selected cases are depicted in Fig \ref{bias} and we observed similar behavior over other ranges of parameters as well. NIBA is known to be problematic in cases with large $\abs{\beta\epsilon}$, often yielding absurd $\abs{\expt{\sigma_z}(t)}>1$. We observe the same behavior here, but also find excellent agreement between GaussSB and HEOM, despite GaussSB and NIBA both only employing second order information $\expt{\sigma_z}^{(2)}(t)$ alone.

\begin{figure}[htb!]
	\centering
	\centering
	(a)\includegraphics[width=3in]{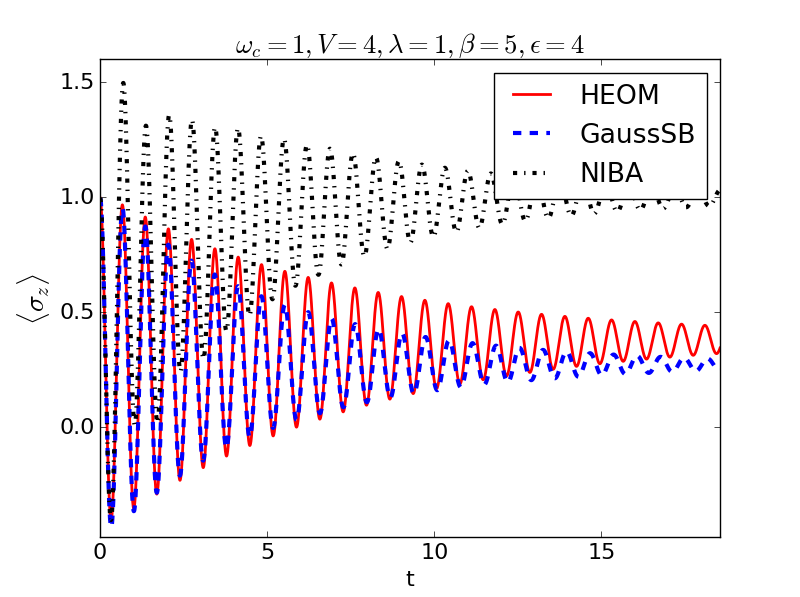}
	\label{fig:nice}
	\vspace{-10pt}
	\centering
	(b)\includegraphics[width=3in]{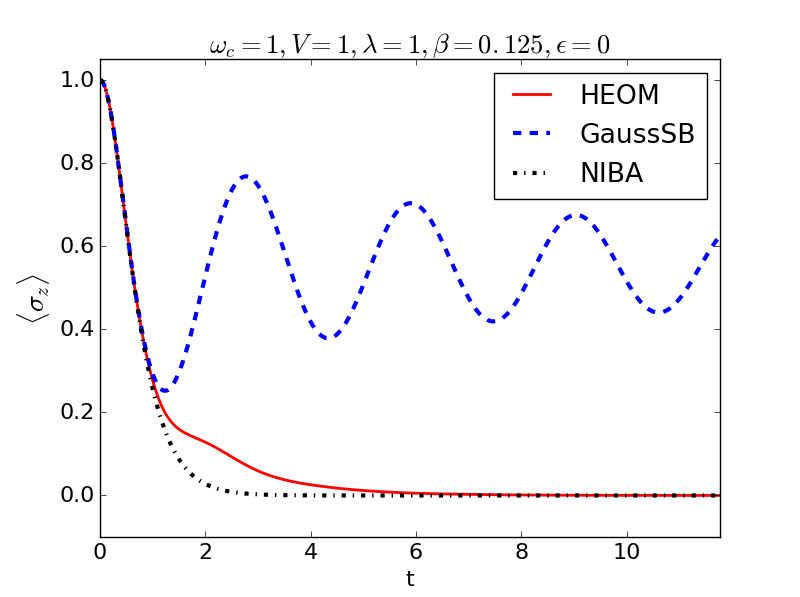}
	\label{fig:nasty}
	\vspace{-10pt}
	\caption{Behavior of GaussSB in fast baths ($\omega_c=1$): (a) The coherent regime, and (b) The dissipative regime.}
	\label{fast}
	\vspace{-8pt}
\end{figure}

\subsection{Behavior for fast baths}
GaussSB cannot be expected to work exactly for fast baths since we explicitly used a small-time approximation taylor series to generate Gaussian traces and sum the perturbation theory series. Numerical tests indicate that this is indeed the case, with GaussSB being qualitatively incorrect in the dissipative regime. We believe that the failure of GaussSB stems from slow decay of coherent oscillations courtesy of the Gaussian approximation (which implicitly assumes slow bath response), as it is the bath response that ultimately causes the transition from oscillatory behavior to dissipation. A related consequence of this is that the formal rate constant associated with GaussSB (obtained from integrating the memory kernel from $t=0$ to $t\to\infty$) is always zero, indicating that GaussSB cannot reproduce correct equilibrium populations or replicate long time exponential decay of the type predicted by Marcus theory and experimentally observed in chemical processes at high temperatures. This represents quite a major obstacle to using pure GaussSB for realistic problems where the dissipative regime is involved.

On the other hand, we do obtain qualitatively correct behavior with GaussSB in regimes where the dynamics is controlled by coherent oscillations as opposed to dissipation (namely cases with large $\beta,V,\epsilon$), but the agreement is not quantitative. We do however note that GaussSB does much better than NIBA in this regime, again despite both of them using only second order information alone. Examples for behavior in both regimes are presented in Fig. \ref{fast}.

\begin{figure*}[ht]
	\vspace{-10pt}
	\begin{minipage}{0.48\textwidth}
		\centering
		(a)\includegraphics[width=3.2in]{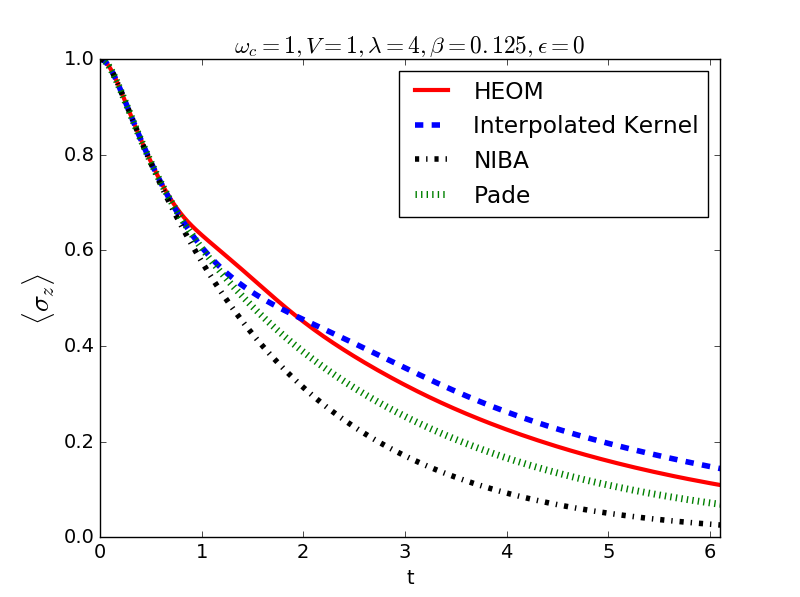}
		\label{fig:fastv1}
	\end{minipage}
	\begin{minipage}{0.48\textwidth}
		\centering
		(b)\includegraphics[width=3.2in]{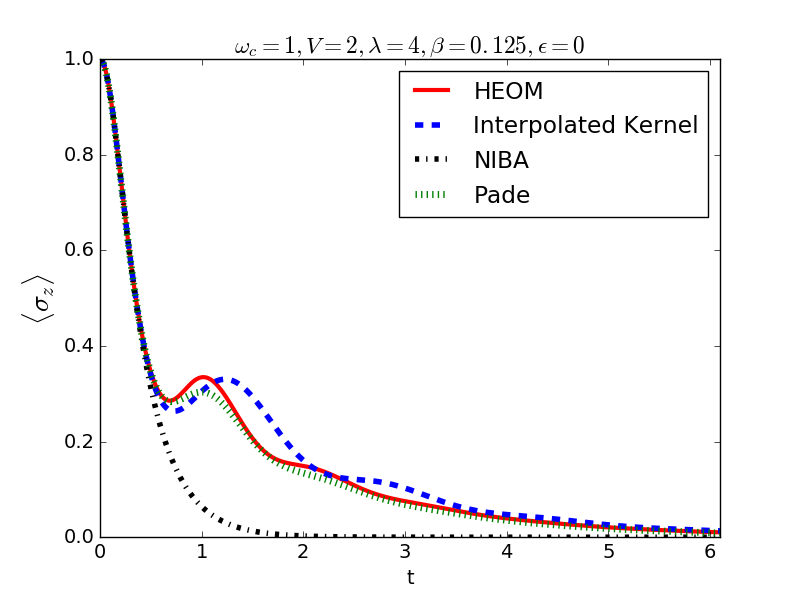}
		\label{fig:fastv2}
	\end{minipage}
	\begin{minipage}{0.48\textwidth}
		\centering
		(c)\includegraphics[width=3.2in]{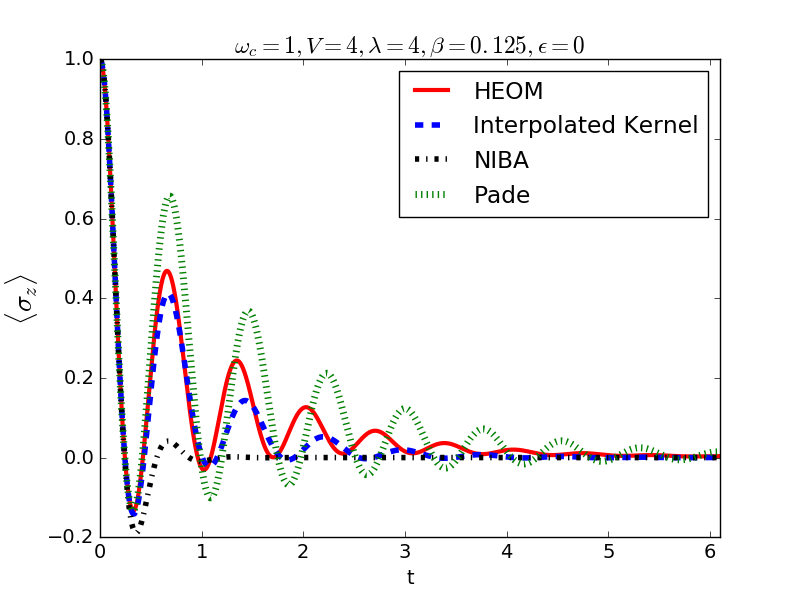}
		\label{fig:fastv4}
	\end{minipage}
	\begin{minipage}{0.48\textwidth}
		\centering
		(d)\includegraphics[width=3.2in]{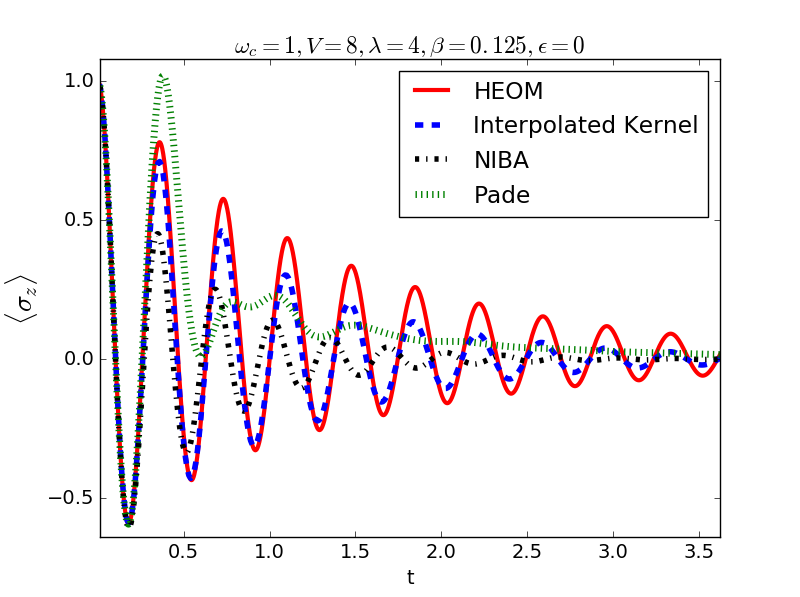}
		\label{fig:fastv8}
	\end{minipage}
	\caption{Comparison of Interpolated kernel populations against HEOM, NIBA and Pad{\'e} resummed kernel correct to fourth order (with $\omega_c=1,\beta=0.125,\lambda=4$ and $\epsilon=0$). The interpolated kernel approach performs better as $V$ increases.}
	\label{nb}
	\vspace{-10pt} 
\end{figure*}

\section{Interpolated Kernels}
Despite the issues related to applicability to fast baths, we know GaussSB is accurate for time-scales shorter than bath response, as evidenced by its accuracy in the slow bath limit.  NIBA is simultaneously known to be accurate at times significantly larger than the bath response as it can reproduce dissipative behavior, indicating that an approach that smoothly ties these two regimes together could potentially be more effective than either. This interpolation scheme should recover pure GaussSB type behavior as $t\to 0$, NIBA type behavior as $t\to \infty$ and behavior intermediate between the two at all times in between.{This is distinct from hybrid quantum-classical methods like MC-TDH\cite{wang2001systematic,thoss2001self}, as it combines exact quantum mechanical results at two limiting cases, as opposed to discriminating between different bath modes by treating only a fraction quantum mechanically.}  

One issue in constructing an interpolating scheme is that GaussSB and NIBA approach population dynamics in completely different ways. NIBA employs a non-Markovian memory kernel to connect $\dot{\expt{\sigma_z}}$ and ${\expt{\sigma_z}}$ while GaussSB directly obtains $\dot{\expt{\sigma_z}}$ via Eqn \ref{gsb1}. The memory kernel typically decays much more rapidly than the population transfer rate\cite{kelly2016generalized}, and thus it seemed more appropriate to connect the memory kernels of GaussSB and NIBA together in order to better utilize the short-time exactness of GaussSB. This led us to numerically invert GaussSB ${\expt{\sigma_z}}$ to obtain kernels $K_{11/22}^{GSB}(t)$ and then smoothly interpolate between these kernels and the exact second order kernels $K^{(2)}_{11/22}(t,t_1)$ to get hybrid kernels that tend to pure second-order at long times and pure GaussSB at short times. Mathematically, this implies:
\begin{align}
K^{I}_{11/22}(t,t_1)&=u(t-t_1)K^{GSB}_{11/22}(t-t_1)\notag\\&\phantom{++}+\left(1-u(t-t_1)\right)K^{(2)}_{11/22}(t,t_1)
\end{align}
where $K^{I}$ is the interpolated kernel and $u(t)$ is an interpolating function with the following properties: 
\begin{enumerate}
	\item $u(t=0)=1$.
	\item $\lim\limits_{t\to \infty} u(t)=0$.
	\item The rate of decay of $u(t)$ (or equivalently, the growth rate of $1-u(t)$) is inversely related to the bath response time. This ensures that GaussSB behavior dies on the timescale of the bath response, since it becomes increasingly inaccurate at long times.
	\item $u(t)$ is well-behaved (i.e., non divergent and continuous, preferably differentiable).
\end{enumerate}
The kernel $K^{I}_{11/22}(t,t_1)$ is thus almost pure GaussSB when the two time indices $t,t_1$ are close, almost pure second order kernel when $t,t_1$ are far and intermediate between the two extremes at intermediate separations. For the equilibrium case, $K^{(2)}_{11/22}(t,t_1)$ is the NIBA kernel, and NIBA behavior is recovered at long times. 
 
There exist many possibilities for $u(t)$, depending on what choices are made regarding bath response times. For the previously employed $J(\omega)=\dfrac{\lambda}{2}\dfrac{\omega\omega_c}{\omega^2+\omega_c^2}$ form, it is at least possible to unambiguously define a timescale by means of $\omega_c$, but such an option is not open for all spectral densities. This leads us to obtain decay rate via $\theta(t)=\dfrac{\displaystyle\int\limits_0^\infty J(\omega)\cos\omega t\coth\dfrac{\beta\omega}{2}\,d\omega}{\displaystyle\int\limits_0^\infty J(\omega)\coth\dfrac{\beta\omega}{2}\,d\omega}$ as $\theta(t)$ corresponds to the normalized energy gap fluctuation autocorrelation function $\dfrac{\mathbf{Re}\left[\expt{\delta\Delta E (t)\delta\Delta E (0)}\right]}{\expt{\delta\Delta E (0)\delta\Delta E (0)}}$, where the numerator is a quantity often computed in MD simulations to obtain $J(\omega)$ and is thus readily available in most cases. This decays smoothly from $1\to 0$ as $t$ runs from $0\to \infty$. Consequently, $\theta(t)$ is a measure of the dissipation rate of initial energy fluctuations and thus gives a timescale for switching to the dissipative regime. 
Furthermore, $\theta(t)=e^{-\omega_c t}$ for Debye spectral densities at small $\beta$ (high temperatures, i.e. the classical limit), indicating that it in fact decays with the same timescale as the bath frequencies.

$\theta(t)$ thus is a good metric for the rate of decay. That is however not the only factor we should consider, as although $\theta(t)$ contains information about bath frequencies as well as some temperature effects, it does not at all account for the energy gap $\epsilon$ which is known to affect population coherence time-scales as well. Furthermore, NIBA is known to be especially terrible for large $\abs{\beta\epsilon}$ (as multiple figures in the preceding section highlight) and this weakness also ought to be considered for building $u(t)$. We therefore propose $u(t)=\abs{\theta(t)}^p$ to be a general form for the interpolating function $u(t)$, where $p$ is very small for $\beta\epsilon\gg1$ (allowing GaussSB to dominate) but reduces to just $1$ for $\epsilon=0$. $p=\mathrm{sech}\beta\epsilon$ was chosen by us as it satsifies the above requirements and corresponds to the $e^{\beta A\left(\epsilon\right)}$, where $A\left(\epsilon\right)=-\dfrac{1}{\beta}\ln \cosh\beta \epsilon$ is the component of the Gibbs-Boguliubov free energy dealing with bias\cite{song2008langevin}. NIBA performance is known to deteriorate as $A(\epsilon)$ increases in magnitude\cite{song2008langevin}, indicating that $p$ can be a decent metric for controlling the GaussSB to NIBA switching rate. The absolute value in $u(t)=\abs{\theta(t)}^p$ is taken since though $\theta(t)\ge 0$ in most cases, a few pathological parameter sets might lead to $\theta(t)<0$, which would make taking non-integer powers problematic. Taking $\abs{\theta(t)}$ solves this issue, though it can lead to non-differentiable $u(t)$ at points where $\theta(t)=0$ changes sign. This is unfortunate, but we feel it is an acceptable compromise due to its rarity and because it does not lead to non-differentiable $\expt{\sigma_z}$ as it only affects the kernel.

For determining the accuracy of this interpolated approach, we considered a few Hamiltonian parameters studied earlier via memory kernel resummation in Ref [\onlinecite{mavros2014resummed}] (all using equilibrium initial conditions) and compared the performance of HEOM, NIBA, Interpolated GaussSB and Pad{\'e} resummed memory kernels\cite{baker1996pade} (obtained from data generated for Ref [\onlinecite{mavros2014resummed}]) correct to fourth order in $V$. The results for cases with $\epsilon=0$ are given in Fig \ref{nb} and it appears that the interpolated kernel approach is comparable to (and for large $V$ is superior to) dynamics obtained from the much more computationally expensive fourth-order resummed kernel approach. It thus appears that the interpolated kernel approach could be a cheap and effective way to incorporate $V^4$ and higher order terms to NIBA at a significantly lower computational cost than numerical evaluation of higher order kernels for resummation. {We also note that in addition to our chosen form of $u(t)$, we experimented with some different forms of $u(t)$ (decaying exponentials and Gaussians in $\omega_ct$) for these cases and obtained roughly similar behavior. This aspect however was not explored in greater detail as other $u(t)$ explored did not easily generalize to non Drude-Lorentz spectral densities where a characteristic $\omega_c$ is not apparent.} 

\begin{figure}[ht]
		(a)\centering
		\includegraphics[width=3.2in]{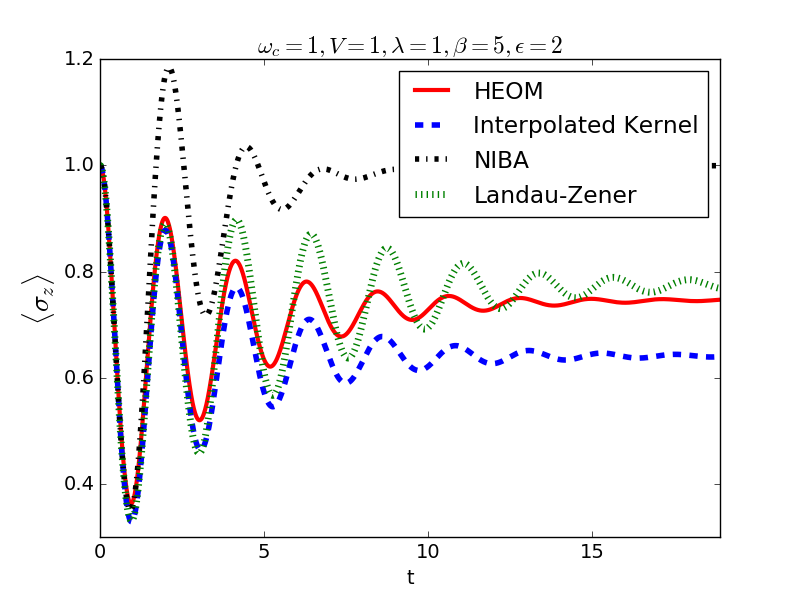}
		\label{fig:bias1}
		\centering
		(b)\includegraphics[width=3.2in]{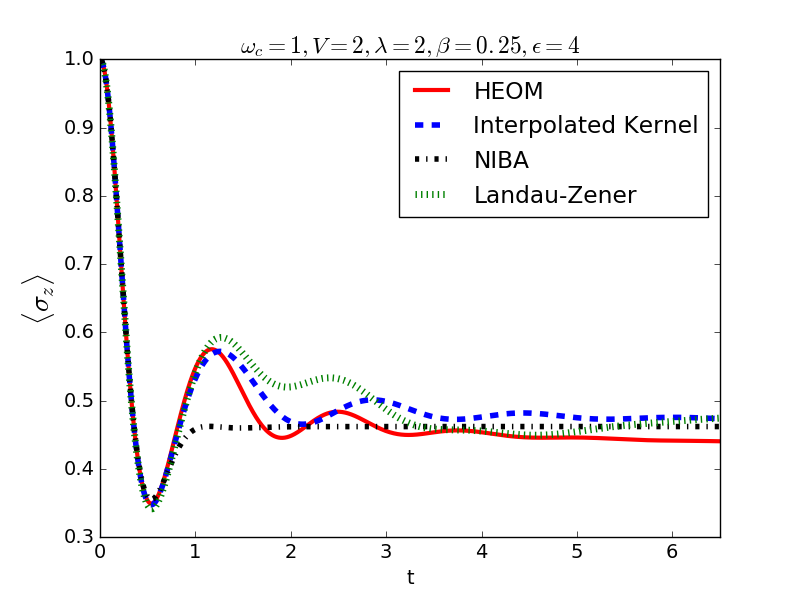}
		\label{fig:bias2}
		\centering
		(c)\includegraphics[width=3.2in]{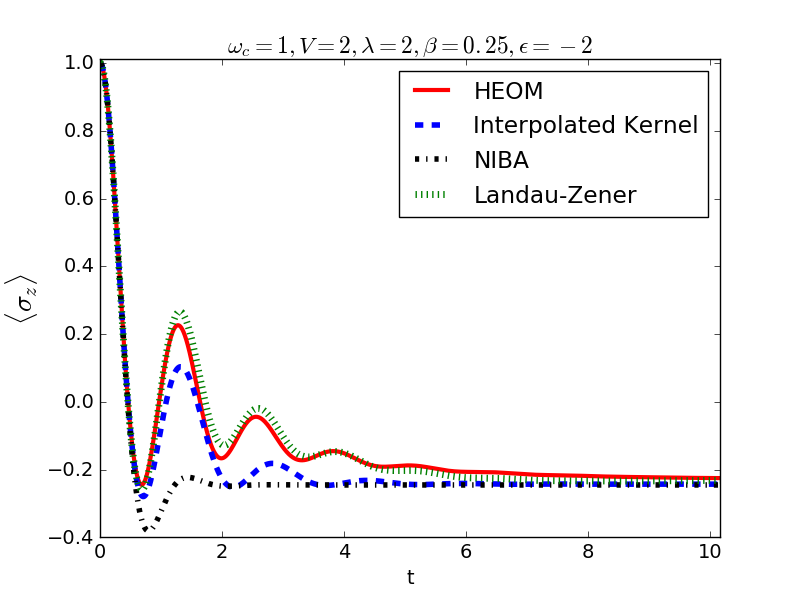}
		\label{fig:bias3}
	\caption{Comparison of interpolated kernel populations against HEOM, NIBA and optimized Landau--Zener resummed kernel correct to fourth order for cases with $\epsilon\ne 0$.}
	\label{bias2}
	\vspace{-12pt}
\end{figure}

Cases with a non-zero $\epsilon$ pose a greater challenge, as we know that neither GaussSB nor NIBA can accurately determine the equilibrium populations in such cases and thus the combined approach is unlikely to succeed in this task either. It has been shown\cite{mavros2014resummed} that it is possible to adjust the resummation scheme to yield kernels correct to fourth order that recover equilibrium populations in a least-squares sense. This indicates that the interpolated kernel in these cases should be expected to perform worse than the optimal fourth order resummed kernel as the latter has more information than the former. Fig \ref{bias2} depicts the comparison of HEOM, NIBA, intepolated kernel and optimized Landau--Zener resummation\cite{sumi1980energy} (found to be superior to Pad{\'e} for cases with bias\cite{mavros2014resummed}) for some systems studied in Ref [\onlinecite{mavros2014resummed}]. This comparison indicates that the interpolated kernel approach gives dynamics of comparable quality to fourth-order resummed kernels, although it tends to perform worse at long times as it does not contain information about $\expt{\sigma_z}(t\to\infty)$ that the resummation technique employs. Nonetheless, it recovers quite reasonable short-time dynamics, with less information and at a fraction of the cost relative to the fourth order methods. {Performance for the difficult cases where exact $\expt{\sigma_z}(t\to\infty)$ is close to $\pm 1$ (such as Fig \ref{fig:bias2}(a)) can potentially be further improved by altering $u(t)$ to ensure recovery of correct equilibrium populations (perhaps via tuning the power parameter $p$).} 


\section{Conclusion and Future Directions}
We have found a relationship between perturbation theory traces of the spin-boson model that allowed us to solve for population dynamics analytically for the case of a Gaussian trace $f_2(t_1,t_2)$. We call this method GaussSB and validate its accuracy by benchmarking it against the highly accurate HEOM method for infinitely slow baths. GaussSB is unfortunately not particularly useful for many problems of chemical interest since $f_2(t_1,t_2)$ is only Gaussian for baths with small frequencies, though it appears to recover the non-dissipative behavior at low temperatures with qualitative accuracy for even fast baths. 

We remedy this by formulating a hybrid kernel that is exact to second order in $V$ and includes higher order correction terms incorporated via GaussSB that decay on the timescale of the bath. This decay is controlled by an interpolating function $u(t)$, in order to prevent higher order GaussSB terms (which become increasingly non-exact at long times) from becoming too influential on timescales larger than bath response. This hybrid approach yields qualitatively accurate dynamics for even fast baths that is comparable in quality to results obtained from using more computationally expensive resummed memory kernels accurate to fourth order. The hybrid kernel method is especially effective in regimes where NIBA normally breaks down (large $\beta,V,\epsilon$), indicating that the interpolation approach was successful in incorporating critical amounts of higher order behavior.

We have thus devised a hybrid method that interpolates between NIBA and GaussSB to give quite accurate dynamics for the spin-boson model for the same computational complexity as NIBA (since it does not need evaluation of expensive triple or higher order integrals). This hybrid approach could thus be used in place of NIBA to explore dynamics across a large parameter space since it has superior performance to NIBA in traditionally problematic regimes and is not observed to ever perform any worse that NIBA, on account of it incorporating higher order effects that NIBA misses. 

It should in principle be possible to generalize this approach to other model Hamiltonians like the linear vibronic coupling (LVC) model\cite{mavhait}, as long as equivalent slow-bath solutions like GaussSB exist. In practice, there are problems stemming from the fact that simple trace factorization relations like Eqn \ref{tracefactor} have not been found (and may simply not exist) for more sophisticated systems.  Directly employing the GaussSB kernel (using the exact $\expt{\dot{\sigma_z}}^{(2)}(t)$ for the model) might be an acceptable compromise for such cases, even though it would sacrifice exactness in the slow-bath limit. The resulting method would resemble a resummation scheme as it would employ the exact second order term to approximate higher order terms, and would furthermore be much cheaper than actually computing fourth order kernels for the problem. We hope to explore the effectiveness of this approach in the future to determine whether it has any promise. 

In the future, we also seek to explore routes to generalize GaussSB to have kernels correct to fourth (or higher) order at all times and see if that yields improved dynamics. The easiest way to achieve this is via a generalization of Eqn \ref{gsb1} of the form
\begin{align}
\dot{\expt{\sigma_z}}(t)=&V^2\dot{\expt{\sigma_z}}^{(2)}(t)\notag\\&-2V^3\displaystyle\int\limits_0^t J_1(2Vf(t_1)t_1)\dot{\expt{\sigma_z}}^{(2)}\left(\sqrt{t^2-t_1^2}\right)\,dt_1
\end{align}
where the function $f(t_1)$ is chosen to ensure $\dot{\expt{\sigma_z}}(t)$ is exact to fourth power in $V$. The kernel resulting from this expression will be exact to fourth order as it contains both second and fourth order information. Generalizing even further, it appears that it might possible to generate other resummation methods employing the general form of Eqn. \ref{gsb1} that are exact in the limit of Gaussian $f_2$. They would still incur significant computational expense in order to have exact fourth (or higher) order behavior, but could potentially prove more effective than pure Pad{\'e} or Landau--Zener resummation and might not even need interpolation on account of the extra information supplied. It may also be possible to recover the correct equilibrium populations with such kernels, and we hope to study the accuracy of such kernels (both with interpolation and without) in the future to determine whether this is indeed the case.

%
%

%

\begin{acknowledgments}
This work was supported by  NSF Grant No. CHE-1058219. D.H. would like to thank the MIT UROP office for support. M.G.M. would also like to thank the NSF GRFP program for funding.
\end{acknowledgments}
\appendix
\section{Time-Dependent Perturbation Theory}
Time-dependent perturbation theory is the necessary first step for any perturbative approaches to quantum dynamics. We treat the diabatic coupling $V$ as the perturbation, which allows us to define the zeroth order Hamiltonian , consequently enabling us to express operator $A(t)$ in the interaction picture as $A_I(t)=e^{i\mathbf{H_0}t}A(t)e^{-i\mathbf{H_0}t}$. The perturbation in the interaction picture is thus $V_I(t)=e^{i\mathbf{H_0}t}\begin{pmatrix}
0&V\\V&0
\end{pmatrix}e^{-i\mathbf{H_0}t}=V\sigma^{x}_I(t)$ ($\sigma^{x}$ being the Pauli matrix in the x direction) and the interaction picture density matrix can be expressed as $\bm{\rho}_I(t)=e^{i\mathbf{H_0}t}\bm{\rho}(t)e^{-i\mathbf{H_0}t}$ (where $\rho(t)$ is the actual time dependent density matrix). The von-Neumann equation for the time-evolution of the interaction-picture density matrix allows us to state that:
\begin{widetext}
\begin{align}
\bm{\rho}_I(t)&=\bm{\rho}_I(0)-i\displaystyle\int\limits_0^t [V_I(t_1),\bm{\rho}_I(0)]\,dt_1-\displaystyle\int\limits_0^t dt_1\displaystyle\int\limits_0^{t_1} [V_I(t_1),[V_I(t_2),\bm{\rho}_I(0)]]\,dt_2\ldots \\
&=\bm{\rho}_I(0)-iV\displaystyle\int\limits_0^t [\sigma^{x}_I(t_1),\bm{\rho}_I(0)]\,dt_1-V^2\displaystyle\int\limits_0^t dt_1\displaystyle\int\limits_0^{t_1} [\sigma^{x}_I(t_1),[\sigma^{x}_I(t_2),\bm{\rho}_I(0)]]\,dt_2\ldots \\
&=\displaystyle\sum\limits_{m=0}^\infty V^m \bm{\rho}^{(m)}_I(t)
\end{align}
\end{widetext}
where $\bm{\rho}^{(m)}_I(t)$ are independent of $V$.
\newline The diabatic populations $p_{1/2}(t)$ can be obtained from the density matrix by tracing out the bath modes and taking the diagonal elements. Mathematically:
\begin{align}
p_1(t)&=\mathbf{Tr}\left[\bm{\rho}_I(t)\dens{1}{1}\right]\\
&=\displaystyle\sum\limits_{m=0}^\infty V^m \mathbf{Tr}_B\left[\bm{\rho}^{(m)}_I(t)\dens{1}{1}\right]\\
p_2(t)&=\mathbf{Tr}\left[\bm{\rho}_I(t)\dens{2}{2}\right]=1-p_1(t)
\end{align}
$p_{1/2}(t)$ are not independent as $p_1(t)+p_2(t)=1$, and thus the population dynamics can be specified with only knowledge of the difference $p_1(t)-p_2(t)$. This is termed as $\expt{\sigma_z}(t)$ as $p_1(t)-p_2(t)=\mathbf{Tr}\left[\bm{\rho}_I(t)\left(\dens{1}{1}-\dens{2}{2}\right)\right]=\mathbf{Tr}\left[\bm{\rho}_I(t)\sigma_z\right]$.

Thus we can express $\expt{\sigma_z}(t)$ as a power series in $V$, with the $m$th order term $\expt{\sigma_z}^{(m)}(t)=\mathbf{Tr}_B\left[\bm{\rho}^{(m)}_I(t)\sigma_z\right]$ for all non-negative integers $m$. Expressing the kernels $K_{11/22}(t)=\displaystyle\sum\limits_{m=0}^\infty V^mK^{(m)}_{11/22}(t)$ as a power series in $V$ and equating terms with same order of $V$ on both sides of Eqns \ref{sm1} and \ref{sm2} allows us to express power series coefficients $K^{(m)}_{11/22}(t)$ in terms of $\expt{\sigma_z}^{(m)}(t)$, which we have already shown can be found from perturbation theory. Going up to some finite order $m$ and finding all $\expt{\sigma_z}^{(k)}(t)$ for $k\le m$ would thus provide us with the correct kernel coefficients $K^{(k)}_{11/22}(t)$, which can be directly employed or be resummed to approximate the exact kernel correct to all orders in $V$. 
\section{Taylor series for $\ln f_2(t_1,t_2)$}
We have:
\begin{widetext}
\begin{align}
f_2(t_1,t_2)&={\mathbf{Tr}\left[e^{i\mathbf{h_1}t_1}e^{-i\mathbf{h_2}t_1}e^{i\mathbf{h_2}t_2}e^{-i\mathbf{h_1}t_2}\rho_B\right]}\\
\dfrac{\partial f_2(t_1,t_2)}{\partial t_1}&=i{\mathbf{Tr}\left[e^{i\mathbf{h_1}t_1}(\mathbf{h_1}-\mathbf{h_2})e^{-i\mathbf{h_2}t_1}e^{i\mathbf{h_2}t_2}e^{-i\mathbf{h_1}t_2}\rho_B\right]}\\
\dfrac{\partial f_2(t_1,t_2)}{\partial t_2}&=-i{\mathbf{Tr}\left[e^{i\mathbf{h_1}t_1}e^{-i\mathbf{h_2}t_1}e^{i\mathbf{h_2}t_2}(\mathbf{h_1}-\mathbf{h_2})e^{-i\mathbf{h_1}t_2}\rho_B\right]}\\
\dfrac{\partial^2 f_2(t_1,t_2)}{\partial t_1^2}&=-{\mathbf{Tr}\left[e^{i\mathbf{h_1}t_1}(\mathbf{h_1}-\mathbf{h_2})^2e^{-i\mathbf{h_2}t_1}e^{i\mathbf{h_2}t_2}e^{-i\mathbf{h_1}t_2}\rho_B\right]}\\
\dfrac{\partial^2 f_2(t_1,t_2)}{\partial t_2^2}&=-{\mathbf{Tr}\left[e^{i\mathbf{h_1}t_1}e^{-i\mathbf{h_2}t_1}e^{i\mathbf{h_2}t_2}(\mathbf{h_1}-\mathbf{h_2})^2e^{-i\mathbf{h_1}t_2}\rho_B\right]}\\
\dfrac{\partial^2 f_2(t_1,t_2)}{\partial t_2\partial t_1}&={\mathbf{Tr}\left[e^{i\mathbf{h_1}t_1}(\mathbf{h_1}-\mathbf{h_2})e^{-i\mathbf{h_2}t_1}e^{i\mathbf{h_2}t_2}(\mathbf{h_1}-\mathbf{h_2})e^{-i\mathbf{h_1}t_2}\rho_B\right]}
\end{align}
\end{widetext}
Since 
\begin{widetext}
\begin{align}
\dfrac{\partial \ln f_2(t_1,t_2)}{\partial t_1}&=\dfrac{1}{f_2(t_1,t_2)}\dfrac{\partial f_2(t_1,t_2)}{\partial t_1}\\
\dfrac{\partial \ln f_2(t_1,t_2)}{\partial t_2}&=\dfrac{1}{f_2(t_1,t_2)}\dfrac{\partial f_2(t_1,t_2)}{\partial t_2}\\
\dfrac{\partial^2 \ln f_2(t_1,t_2)}{\partial t_1^2}&=\dfrac{1}{f_2(t_1,t_2)}\dfrac{\partial^2 f_2(t_1,t_2)}{\partial t_1^2}-\left(\dfrac{1}{f_2(t_1,t_2)}\right)^2\left(\dfrac{\partial  f_2(t_1,t_2)}{\partial t_1}\right)^2\\
\dfrac{\partial^2 \ln f_2(t_1,t_2)}{\partial t_2^2}&=\dfrac{1}{f_2(t_1,t_2)}\dfrac{\partial^2 f_2(t_1,t_2)}{\partial t_2^2}-\left(\dfrac{1}{f_2(t_1,t_2)}\right)^2\left(\dfrac{\partial  f_2(t_1,t_2)}{\partial t_2}\right)^2\\
\dfrac{\partial^2 \ln f_2(t_1,t_2)}{\partial t_1\partial t_2}&=-\left(\dfrac{1}{f_2(t_1,t_2)}\right)^2\dfrac{\partial f_2(t_1,t_2)}{\partial t_1}\dfrac{\partial f_2(t_1,t_2)}{\partial t_2}+f_2(t_1,t_2)\dfrac{\partial^2 f_2(t_1,t_2)}{\partial t_1\partial t_2}
\end{align}
\end{widetext}
Expanding $\ln f_2(t_1,t_2)$ as a Taylor series about $t_1=t_2=0$ gives:
\begin{widetext}
\begin{align}
\ln f_2(t_1,t_2)=0+i(t_1-t_2){\mathbf{Tr}\left[(\mathbf{h_1}-\mathbf{h_2})\rho_B\right]}-\dfrac{1}{2}\left({\mathbf{Tr}\left[(\mathbf{h_1}-\mathbf{h_2})^2\rho_B\right]}-\left({\mathbf{Tr}\left[(\mathbf{h_1}-\mathbf{h_2})\rho_B\right]}\right)^2\right)\left(t_1-t_2\right)^2\ldots 
\end{align}
Higher order terms can be neglected if:
\begin{align}
\dfrac{\partial^3 \ln f_2(t_1,t_2)}{\partial t_1^3}&=i\left(3{\mathbf{Tr}\left[(\mathbf{h_1}-\mathbf{h_2})^2\rho_B\right]}{\mathbf{Tr}\left[(\mathbf{h_1}-\mathbf{h_2})\rho_B\right]}-{\mathbf{Tr}\left[(\mathbf{h_1}-\mathbf{h_2})^3\rho_B\right]}-2\left({\mathbf{Tr}\left[(\mathbf{h_1}-\mathbf{h_2})\rho_B\right]}\right)^3\right)
\end{align}
\end{widetext}
etc are small, which is the case if the bath frequencies $\{\omega_i\}$ are small. As an example, the combined third order term for equilibrium $\rho_{B}=\dfrac{e^{-\beta \mathbf{h_1}}}{\mathbf{Tr}\left[e^{-\beta \mathbf{h_1}}\right]}$ is $-\dfrac{2i}{3\pi}(t_1-t_2)^3\displaystyle\int\limits_0^\infty \omega J(\omega)\,d\omega$, which is guaranteed to be small if $J(\omega)$ is only considerable for small $\omega$. Similarly, the fourth order term for this case is:
$
\dfrac{(t_1-t_2)^4}{6\pi}\displaystyle\int\limits_0^\infty \omega^2 J(\omega)\coth \dfrac{\beta\omega}{2}\,d\omega
$, which will be small if $J(\omega)$ is only large for small $\omega$.
\section{Derivation of {Slow Bath} Solution}
Integrating Gaussian $f_{2n}$ analytically yields a relation between population growth terms $\dot{\expt{\sigma_z}}^{(2n)}$:
\begin{align}
\dot{\expt{\sigma_z}}^{(2n)}(t)&=-\dfrac{2}{n-1}\displaystyle\int\limits_0^t t_1\dot{\expt{\sigma_z}}^{(2n-2)}(t_1)\,dt_1
\end{align}
for all integers $n>1$.
This form is more convenient to handle in Laplace space where we have:
\begin{align}
\mathcal{L}\left[\dot{\expt{\sigma_z}}^{(2n)}\right](s)&=\dfrac{2}{n-1}\dfrac{1}{s} \dfrac{d}{ds}\mathcal{L}\left[\dot{\expt{\sigma_z}}^{(2n-2)}\right](s)\label{lap}\\
&=\dfrac{4}{(n-1)(n-2)}\dfrac{1}{s} \dfrac{d}{ds}\dfrac{1}{s} \dfrac{d}{ds}\mathcal{L}\left[\dot{\expt{\sigma_z}}^{(2n-4)}\right](s)\\
&=\dfrac{1}{(n-1)!}\left(\dfrac{2}{s} \dfrac{d}{ds}\right)^{n-1}\mathcal{L}\left[\dot{\expt{\sigma_z}}^{(2)}\right](s)
\end{align}
by repeated application of Eqn \ref{lap}. This permits us to sum the infinite power series for $\dot{\expt{\sigma_z}}$ in Laplace space, as we have:
\begin{align}
\mathcal{L}\left[\dot{\expt{\sigma_z}}\right](s)&=\mathcal{L}\left[\displaystyle\sum\limits_{n=1}^\infty V^{2n}\dot{\expt{\sigma_z}}^{(2n)}\right](s)\\
&=\displaystyle\sum\limits_{n=1}^\infty\dfrac{V^{2n}}{(n-1)!}\left(\dfrac{2}{s} \dfrac{d}{ds}\right)^{n-1}\mathcal{L}\left[\dot{\expt{\sigma_z}}^{(2)}\right](s)\\
&=V^2\exp\left(\dfrac{2V^2}{s} \dfrac{d}{ds}\right)\mathcal{L}\left[\dot{\expt{\sigma_z}}^{(2)}\right](s)\\
&=V^2\exp\left(4V^2\dfrac{d}{ds^2}\right)\mathcal{L}\left[\dot{\expt{\sigma_z}}^{(2)}\right](s)\\
&=V^2\mathcal{L}\left[\dot{\expt{\sigma_z}}^{(2)}\right](\sqrt{s^2+4V^2})
\end{align}
Returning to the time domain, we have via known inverse Laplace transforms\cite{bateman1954tables}:
\begin{align}
\dot{\expt{\sigma_z}}(t)&=V^2\dot{\expt{\sigma_z}}^{(2)}(t)-2V^3\displaystyle\int\limits_0^t J_1(2Vt_1)\dot{\expt{\sigma_z}}^{(2)}\left(\sqrt{t^2-t_1^2}\right)\,dt_1\label{gsb1a}\\
\dot{\expt{\sigma_z}}^{(2)}(t)&=-4\displaystyle\int\limits_0^t\mathbf{Re}\left[f_2(t,t_1)\right]\,dt_1\\&=-\mathbf{Re}\left[\frac{2\sqrt{\pi } e^{-\frac{b^2}{4 a}} \text{erf}\left(\frac{2 a t+ib}{2 \sqrt{a}}\right)}{ \sqrt{a}}\right]\label{gsb2a}
\end{align}
where $a\ne 0$. $J_1$ is a Bessel function of the first kind, giving us an analytical solution to the spin-boson model for all baths with trace $f_2$ Gaussian in time. 

\section{GaussSB in the limit of Rabi Oscillations}
If we consider the trivial case of $J(\omega)=0$, we obtain $f_2(t_1,t_2)=e^{i\epsilon(t_1-t_2)}$, which is a complex exponential and not actually a Gaussian. Thus $a=0$ and we can't use Eqn. \ref{gsb2} directly, but it is easy to use Eqn. \ref{gsb1a} and find:
\begin{align}
\dot{\expt{\sigma_z}}^{(2)}(t)&=-4\displaystyle\int\limits_0^t\mathbf{Re}\left[f_2(t,t_1)\right]\,dt_1=-\dfrac{4\sin \epsilon t}{\epsilon}
\end{align}
\begin{align}
\dot{\expt{\sigma_z}}(t)&=-\frac{4 \sin \left(t \sqrt{4 V^2+\epsilon ^2}\right)}{\sqrt{4 V^2+\epsilon ^2}}
\\
\implies \expt{\sigma_z}(t)&=\frac{4 V^2 \cos \left(t \sqrt{4 V^2+\epsilon ^2}\right)+\epsilon ^2}{4 V^2+\epsilon ^2}
\end{align}
which is the exact result that can be directly calculated for this trivial example, showing that it correctly recovers the Rabi limit. 
\section{Kernels from Populations}
GaussSB gives $\expt{\sigma_z}(t)$ directly and does not employ any memory kernels. Nonetheless, it would be preferable to interpolate between time non-local memory kernels that population growth rates to fully exploit the the complete short-time information contained in GaussSB, compelling us to determine ways to invert the GaussSB populations $\expt{\sigma_z}(t)$ to yield kernels.
\newline We begin from the one time-index version of the Sprapaglione-Mukamel formalism\cite{muk} as there is no way (or reason) to introduce a second time index whose behavior we cannot solely recover from the univariate $\expt{\sigma_z}(t)$. Specifically, we use a variant of Eqn \ref{sm3} where:
{\small\begin{align}
\dfrac{d\expt{\sigma_z}}{dt}&=-\displaystyle\int\limits_{0}^t\left(K_{-}(t-t_1)+K_{+}(t-t_1)\expt{\sigma_z}(t_1)\right)\,dt_1\\
&=-\displaystyle\int\limits_{0}^tK_{+}(t_1)\expt{\sigma_z}(t-t_1)dt_1-\displaystyle\int\limits_{0}^tK_{-}(t_1)dt_1\label{test}
\end{align}}
Differentiating Eqn. \ref{test} with respect to time, we obtain:
{\small\begin{align}
\ddot{\expt{\sigma_z}}(t)&=-\displaystyle\int\limits_{0}^tK_{+}(t_1)\dot{\expt{\sigma_z}}(t-t_1)dt_1-K_{-}(t)-K_{+}(t)\expt{\sigma_z}(0)\label{lapa}
\end{align}}
Let us consider two initial conditions: $\expt{\sigma_z}(0)=\pm 1$ and call the corresponding $\expt{\sigma_z}(t)$ as $\expt{\sigma_z}_\pm(t)$. Then we have:

{\small\begin{align}
&\ddot{\expt{\sigma_z}}_{\pm}(t)=-\displaystyle\int\limits_{0}^tK_{+}(t_1)\dot{\expt{\sigma_z}}_+(t-t_1)dt_1-K_{-}(t)\mp K_{+}(t)\label{Km}\\
&\implies \ddot{\expt{\sigma_z}}_+(t)-\ddot{\expt{\sigma_z}}_-(t)=-\displaystyle\int\limits_{0}^tK_{+}(t_1)\left(\dot{\expt{\sigma_z}}_+(t-t_1)\right.\notag\\&\phantom{uygawfugagwhejsrhdr}\left.-\dot{\expt{\sigma_z}}_-(t-t_1)\right)dt_1-2K_{+}(t)\label{K}
\end{align}
}
GaussSB gives us $\dot{\expt{\sigma_z}}_\pm(t-t_1)$ from which we can obtain the second derivatives $\ddot{\expt{\sigma_z}}_{\pm}(t-t_1)$ via direct differentiation of Eqn \ref{gsb1} or finite differences (which was the route we took). Then we can solve Eqn. \ref{K} numerically on a grid for $K_{+}(t)$ (by treating the integral as a finite sum via trapezoid rule and solving the resulting system of linear equations for instance, as we have done: but other approaches like Archimedes summation or other Newton-Coates approaches should be equally valid), making use of the fact that $K_{+}(0)=4V^2$ since only second order terms matter then. Finally, back substitution of $K_+(t)$ in Eqn \ref{Km} is sufficient to recover $K_-(t)$, yielding the kernels. 
\section{Kernels in Laplace Space}
Eqn. \ref{K} offers an interesting alternative approach for accessing kernels via Laplace space. Taking Laplace transforms on both sides, we obtain:
\begin{widetext}
\begin{align}
s\left(\mathcal{L}\left[\dot{\expt{\sigma_z}}_+\right](s)-\mathcal{L}\left[\dot{\expt{\sigma_z}}_-\right](s)\right)&=-\mathcal{L}\left[K_{+}\right](s)\left(\mathcal{L}\left[\dot{\expt{\sigma_z}}_+\right](s)-\mathcal{L}\left[\dot{\expt{\sigma_z}}_-\right](s)\right)-2\mathcal{L}\left[K_{+}\right](s)\\
\implies \mathcal{L}\left[K_{+}\right](s)&=-\dfrac{s\left(\mathcal{L}\left[\dot{\expt{\sigma_z}}_+\right](s)-\mathcal{L}\left[\dot{\expt{\sigma_z}}_-\right](s)\right)}{2+\mathcal{L}\left[\dot{\expt{\sigma_z}}_+\right](s)-\mathcal{L}\left[\dot{\expt{\sigma_z}}_-\right](s)}\\
&=-\dfrac{sV^2\left(\mathcal{L}\left[\dot{\expt{\sigma_z}}^{(2)}_+\right](\sqrt{s^2+4V^2})-\mathcal{L}\left[\dot{\expt{\sigma_z}}^{(2)}_-\right](\sqrt{s^2+4V^2})\right)}{2+V^2\mathcal{L}\left[\dot{\expt{\sigma_z}}^{(2)}_+\right](\sqrt{s^2+4V^2})-V^2\mathcal{L}\left[\dot{\expt{\sigma_z}}^{(2)}_-\right](\sqrt{s^2+4V^2})}
\end{align}
\end{widetext}
We can determine $\mathcal{L}\left[\dot{\expt{\sigma_z}}^{(2)}_\pm\right](s)$ via simple Laplace transform from time-domain and then use it to find $\mathcal{L}\left[K_{+}\right](s)$ (and similarly $\mathcal{L}\left[K_{-}\right](s)$). However, inverse Laplace transforms are numerically unstable and thus using this route to obtain time domain kernels is suboptimal.

\bibliography{references}

\end{document}